\documentclass[%
 reprint,
 amsmath,amssymb,
 aps,
]{revtex4-2}

\usepackage{graphicx}
\usepackage{bm}
\usepackage{xcolor}
\usepackage{booktabs} 
\usepackage{siunitx}
\usepackage{amsfonts, amssymb, amsmath}
\usepackage{upgreek}
\usepackage{physics} 

\newcommand{\thor}{$^{229}$Th}
\newcommand{\thorm}{$^\mathrm{229m}$Th}
\newcommand{\caf}{CaF$_2$}

\newcommand{\Vzz}{$V_{\rm zz}$}
\newcommand{\Vxx}{$V_{\rm xx}$}
\newcommand{\Vyy}{$V_{\rm yy}$}
\newcommand{\Qg}{$Q_{\rm g}$}
\newcommand{\Qe}{$Q_{\rm e}$}

\newcommand{\gray}[1]{\textcolor{gray}{#1}}

\usepackage{graphicx}
\usepackage{dcolumn}
\usepackage{bm}

\begin{document}

\preprint{APS/123-QED}

\title{Temperature sensitivity of a Thorium-229 solid-state nuclear clock}
\author{Jacob S. Higgins, Tian Ooi, Jack F. Doyle, Chuankun Zhang, Jun Ye}
\email{email: ye@jila.colorado.edu}
\affiliation{JILA, NIST and University of Colorado, Department of Physics, University of Colorado, Boulder, CO 80309}

\author{Kjeld Beeks}
\author{Tomas Sikorsky}
\author{Thorsten Schumm}
\affiliation{Vienna Center for Quantum Science and Technology, Atominstitut, TU Wien, 1020 Vienna, Austria}

\date{\today}

\begin{abstract}
Quantum state-resolved spectroscopy of the low energy thorium-229 nuclear transition was recently achieved. The five allowed transitions within the electric quadrupole structure were measured to the kilohertz level in a calcium fluoride host crystal, opening many new areas of research using nuclear clocks. Central to the performance of solid-state clock operation is an understanding of systematic shifts such as the temperature dependence of the clock transitions. In this work, we measure the four strongest transitions of thorium-229 in the same crystal at three temperature values: 150\,K, 229\,K, and 293\,K. We find shifts of the unsplit frequency and the electric quadrupole splittings, corresponding to decreases in the electron density, electric field gradient, and field gradient asymmetry at the nucleus as temperature increases. The \textit{m} = $\pm$5/2 → $\pm$3/2 line shifts only 62(6)\,kHz over the temperature range, i.e., approximately 0.4\,kHz/K, representing a promising candidate for a future solid-state optical clock. Achieving 10$^{-18}$ precision requires crystal temperature stability of 5\,$\upmu$K.
\end{abstract}

\maketitle

\noindent \textit{Introduction.} Nuclear clocks promise a new regime of scientific measurement capability. This includes searches for ultralight dark matter candidates through variations of fundamental constants~\cite{Flambaum2006,Peik2021,Zhang2024,Beeks2024}, solid-state optical clocks for field applications~\cite{Rellergert2010, Kazakov2012, Zhang2024b}, and more precise clock operation in trapped ions~\cite{Campbell2011,Thielking2018, Yamaguchi2024}. Current optical atomic clocks based on electronic transitions utilize quantum state engineering for precise measurement of high quality factor resonances and detailed characterization of systematic effects to account for or reduce shifts and broadenings \cite{Bloom2014,McGrew2018,Bothwell2022}. Examples in these cold atom systems include blackbody radiation shifts, light shifts, and Zeeman splittings and shifts~\cite{Aeppli2024}. The systematic shifts and splittings of nuclear transitions must be characterized to achieve the level of precision of optical atomic clocks. 

\begin{figure*}[th!]
    \includegraphics[width=14cm]{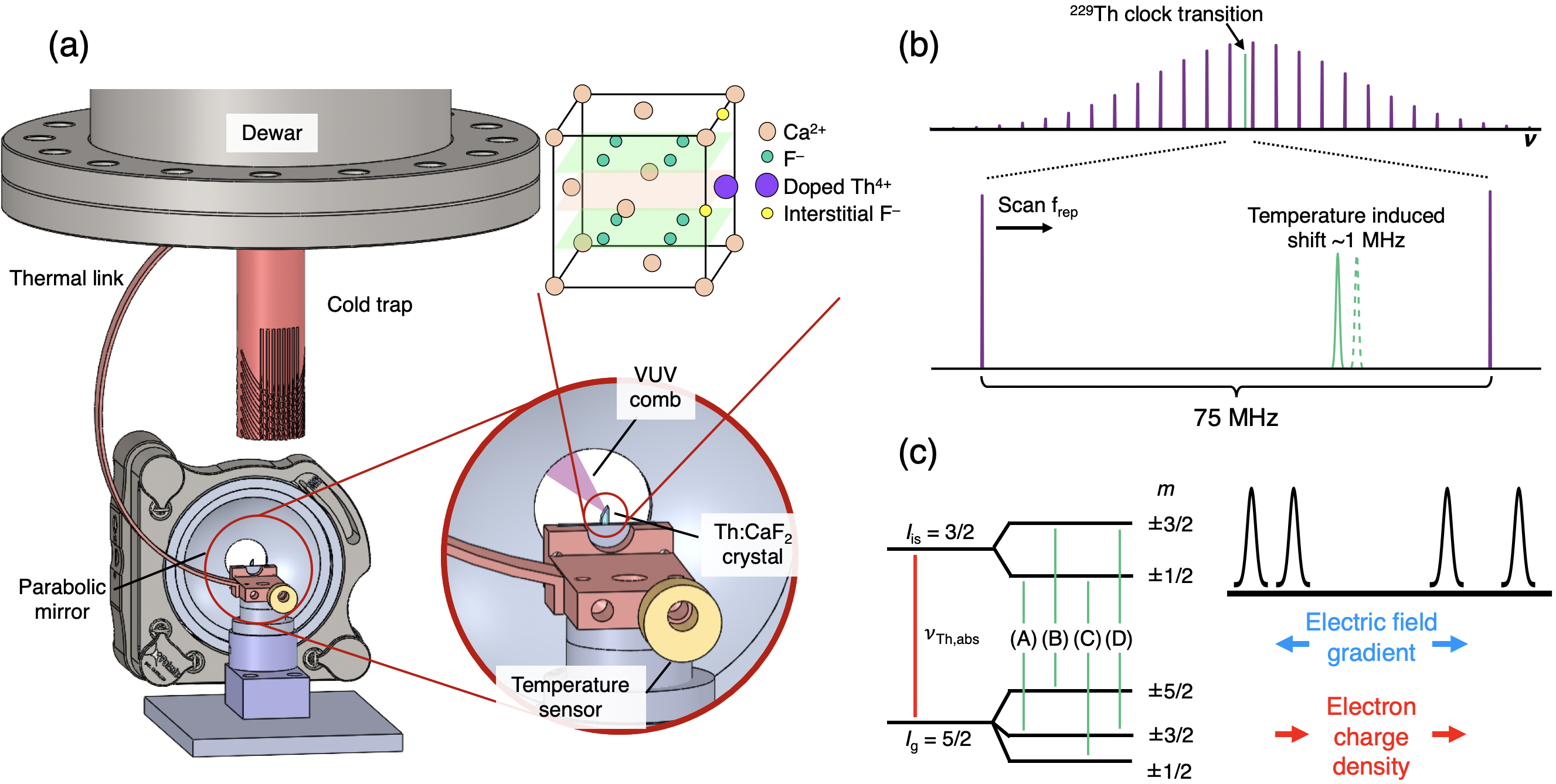}
    \caption{
        (a) Schematic of the crystal mount and thermal control. The thorium-doped calcium fluoride crystal is mounted on a baseplate in vacuum, which is connected to an external dewar via a copper thermal link. The temperature is continually monitored using a sensor attached to the mounting plate. The dewar is in contact with a cold trap which condenses molecules to reduce vacuum pressure. The dewar is filled with liquid nitrogen for measurements at 150\,K, a methanol/dry ice mixture for measurements at 229\,K, and is unfilled for measurements at 293\,K. A parabolic mirror collimates fluorescent photons, which are then counted downstream after spectral filtering. Shown in the inset is the crystal structure of calcium fluoride, where one calcium ion is replaced with doped Th$^{4+}$, and two F$^{–}$ interstitials are added for charge compensation. (b) Frequency comb spectroscopy of thorium. The comb mode number exciting the transition was determined in previous work (see \textit{Methods})~\cite{Zhang2024}. The frequency of each spectral line is determined by scanning \textit{f}$_{\mathrm{rep}}$ and fitting to the line center. (c) Electric quadrupole structure of \thor, giving rise to magnetic dipole allowed transitions (see Table~\ref{tab:freqs}). Increasing the electric field gradient changes the electric quadrupole splitting, causing the lines to spread outward; increasing the electron density at the nucleus induces an isomeric shift of all lines (see Main Text). Not shown is the \textit{m} = $\pm$1/2 → $\pm$3/2 transition, whose intensity is an order of magnitude weaker than the other four lines and is not measured in this study.
        }
    \label{fig:setup}
\end{figure*}
     
X-ray photon sources have coherently excited nuclear states~\cite{Haber2017,Heeg2021}, including a potential nuclear clock transition for $^{45}$Sc at 12.4\,keV~\cite{Shvyd2023}. Excitation of the 76\,eV clock transition in $^{235}$U using a resonant electron bridge process has been proposed~\cite{Berengut2018}. The thorium-229 nucleus has a uniquely low energy nuclear transition (\thorm) at 8.4\,eV (2020\,THz), accessible with vacuum ultraviolet (VUV) lasers~\cite{vonderWense2016,Seiferle2019,Kraemer2023,Hiraki2024}. The last two years have seen rapid progress in the development of thorium-based clocks, with the first nuclear fluorescence reported in 2023~\cite{Kraemer2023} and the first~\cite{Tiedau2024} and second~\cite{Elwell2024} instances of direct laser excitation occurring in 2024. These experiments constrained the transition uncertainty to the ten gigahertz level. Later in 2024, a coherent, high spectral-resolution VUV frequency comb~\cite{Cingoz2012,Zhang2022} was used to directly resolve individual quantum states of the nuclear transition in a calcium fluoride (\caf) crystal, and connected the absolute frequency to the $^{87}$Sr optical atomic clock as an absolute frequency reference~\cite{Zhang2024}. This new nuclear frequency standard placed the measurement uncertainty of the \thorm{} transition at the kilohertz level. The resolved nuclear electric quadrupole structure allowed the determination of the ratio of the ground state's electric quadrupole moment to that of the isomeric state to be \Qe /\Qg =0.57003(1).  

The thorium nuclear transition is an excellent candidate for a next-generation optical frequency standard~\cite{Tamm2003,Rellergert2010,Campbell2011,Beeks2021}. The \thorm{} lifetime in CaF$_{2}$ was measured to be 641(4)\,s, placing the projected oscillator quality factor at potentially the 10$^{19}$ level \cite{Kraemer2023, Tiedau2024, Elwell2024, Hiraki2024, Yamaguchi2024, Zhang2024}. In addition, the small multipole moments of nuclei compared to electron orbits make the nuclear transition much less sensitive to external perturbations. It has been suggested that the transition will have an enhanced sensitivity to variations of fundamental constants such as the fine structure constant $\alpha$~\cite{Flambaum2006,Rellergert2010}. Using the quadrupole ratio measured in Ref.~\cite{Zhang2024}, the enhancement factor was estimated to be three to four orders of magnitude higher than that in current optical atomic clocks~\cite{Beeks2024,Caputo2024}. 

Besides the application in metrology, the thorium nuclear transition can also be used to precisely probe material properties such as the local electron density. Mössbauer spectroscopy uses nuclear transitions to study the chemical environment surrounding the nucleus, including oxidation state, local magnetic field couplings, and electron distribution~\cite{Gutlich2012,Bianchi2021}. It is useful for many applications such as spin crossover in organometallic compounds~\cite{Kuzmann2021}, bioinorganic systems~\cite{Kamnev2021} including metalloproteins~\cite{Kamnev2017}, nanomaterials~\cite{Alenkina2022}, and temperature dependent studies of solids~\cite{Weber2014}. There are over 45 active Mössbauer elements used today~\cite{Bianchi2021}, including thorium-232~\cite{Hershkowitz1968}. Mössbauer spectroscopy is typically conducted using gamma ray sources or synchrotron radiation. The thorium-229 nucleus is the first and possibly only Mössbauer active transition accessible by current tabletop laser sources.

In this work, we measure the \thorm{} transition at multiple temperature (\textit{T}) values and extract the \textit{T}-dependent electric quadrupole splittings and the unsplit, or isomeric shift. Both the electric field gradient (EFG) and the unsplit transition frequency show clear changes with respect to temperature. These indicate that as \textit{T} increases, the electron density and EFG at the nucleus both decrease (see discussion below). One line, the \textit{m} = $\pm$5/2 → $\pm$3/2 transition, is over an order of magnitude less sensitive to temperature change due to partial cancellation of the splitting and the isomeric shift, making it suitable for future stable clock operation.

\noindent \textit{Absolute line frequencies.} Our experimental setup is shown in Fig.~\ref{fig:setup}. The \thor{} ground state has a nuclear spin of I = 5/2, while the isomeric state \thorm{} has nuclear spin I = 3/2. This gives rise to five allowed magnetic dipole transitions given by the selection rule $\Delta m$ = 0 or $\pm$1, where $m$ is the angular momentum projection onto the \textit{z}-axis. We scanned four of the five allowed transitions within the electric quadrupole structure (see \textit{Methods} for details). The absolute frequency for each spectroscopic line at the three temperatures is shown in Table~\ref{tab:freqs}. The fifth line corresponding to \textit{m}~= $\pm$1/2 → $\pm$3/2 was not scanned due to its relatively weaker transition strength~\cite{Kazakov2012} but was obtained using the sum rule described in Ref.~\cite{Zhang2024}. This line is shown in the table in gray.  

\begin{figure*}[th!]
    \includegraphics[width=12cm]{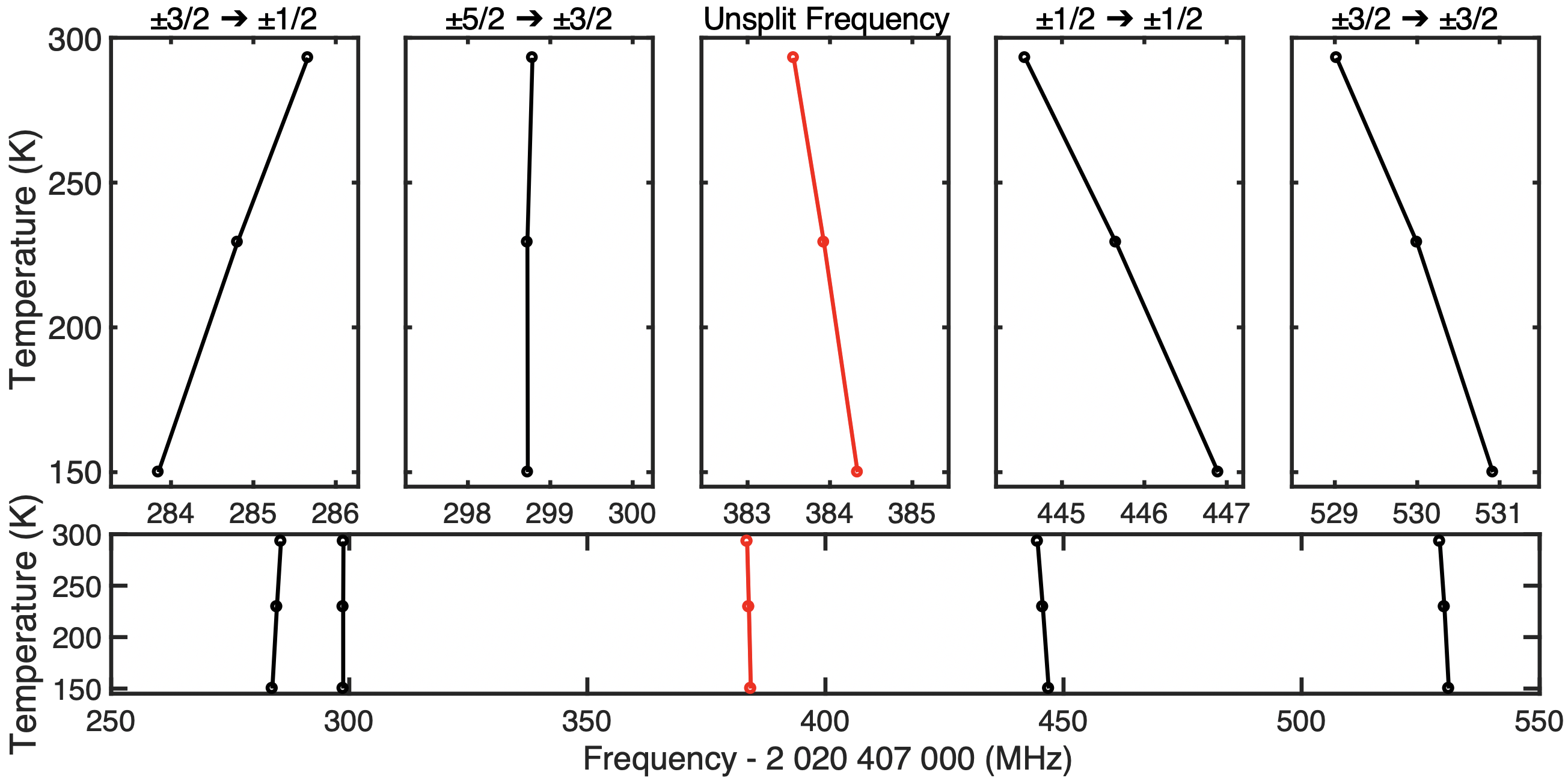}
    \caption{
    Frequencies of the four prominent electric quadrupole lines plotted against temperature. Shown at the top are zoomed views of the full spectrum, plotted at the bottom. The frequency axes of the top plots all have a 3\,MHz range. The red line shows the isomeric, or unsplit shift, representing the I = 5/2 → 3/2 transition. The unsplit frequency decreases with increasing temperature, indicating that electron density at the nucleus decreases. The four lines show reduced quadrupole splitting when temperature is increased, indicating a smaller electric field gradient. The \textit{m} = $\pm$5/2 → $\pm$3/2 transition is the least temperature sensitive transition due to a partial cancellation of these two effects.
    }
    \label{fig:temp_plot}
\end{figure*}

Fig.~\ref{fig:temp_plot} shows the measured transition energies plotted against \textit{T}, shown in black. The top section is a series of zoomed in plots with the frequency axis spanning 3\,MHz for each line to show the relative \textit{T}-dependence of the four features. Plotted in red is the unsplit transition frequency, calculated using the formula 
\begin{equation}
\nu_{\rm Th,abs}=\frac{1}{6}(\nu_{\frac{3}{2} \rightarrow \frac{1}{2}}+2\nu_{\frac{5}{2} \rightarrow \frac{3}{2}}+2\nu_{\frac{1}{2} \rightarrow \frac{1}{2}}+\nu_{\frac{3}{2} \rightarrow \frac{3}{2}}).
    \label{eq:unsplitFreq}
\end{equation}
\noindent This frequency, free from quadrupole splittings, is obtained by taking a proper average of the four measured transitions; this calculation is valid even under asymmetric splitting (nonzero $\eta$, see Eq.~\ref{eq:quadrupole}). The unsplit frequency is reduced by 776\,kHz at room temperature relative to 150\,K, indicating a change in electronic charge density at the nucleus. In addition, the peak splittings decrease as \textit{T} is increased, due to the change in EFG. Three of the peaks exhibit a clear shift on the order of a few MHz. One peak, the \textit{m} = $\pm$5/2 → $\pm$3/2 transition, shows roughly one order of magnitude smaller sensitivity to temperature, with a shift of only 62(6)\,kHz over the 143\,K range of temperature changes.

\setlength{\tabcolsep}{0.8em} 
{\renewcommand{\arraystretch}{0.5}
\begin{table*}[t]
\caption{\label{tab:freqs}%
Measured (black) and calculated (gray, starred) frequencies at three different temperatures, corresponding to Fig \ref{fig:temp_plot}. The EFG-free field is obtained using Eq. \ref{eq:unsplitFreq}.}
\begin{tabular}{|c|ccc|}
\hline
Transition                         & \begin{tabular}[c]{@{}c@{}}Abs freq (MHz)\\ 150(1)\,K\end{tabular} & \begin{tabular}[c]{@{}c@{}}Abs freq (MHz) \\ 229(1)\,K\end{tabular} & \begin{tabular}[c]{@{}c@{}}Abs freq (MHz)\\ 293(1)\,K\end{tabular} \\ \hline
(A)    3/2 $\rightarrow$ 1/2 & 2020 407 283.847(4)                                                   & 2020 407 284.808(8)                                                         & 2020 407 285.662(5)                                                        \\
(B)    5/2 $\rightarrow$ 3/2 & 2020 407 298.727(4)                                                   & 2020 407 298.722(3)                                                         & 2020 407 298.784(5)                                                        \\
(C)   1/2 $\rightarrow$ 1/2  & 2020 407 446.895(4)                                                   & 2020 407 445.654(5)                                                         & 2020 407 444.551(3)                                                        \\
(D)    3/2 $\rightarrow$ 3/2 & 2020 407 530.918(4)                                                   & 2020 407 529.996(5)                                                         & 2020 407 529.021(3)       
\\
(E) 1/2 $\rightarrow$ 3/2 & 2020 407 693.98(2) &  \gray{2020 407 690.84(1)*}  &  \gray{2020 407 687.910(7)*}     \\ \hline

\gray{Unsplit frequency}                    & \gray{2020 407 384.335(2)*}                                                   & \gray{2020 407 383.926(3)*}                                                         & \gray{2020 407 383.559(2)*}                                                        \\ \hline
\end{tabular}
\end{table*}
}

\setlength{\tabcolsep}{0.8em} 
{\renewcommand{\arraystretch}{1.2}

\noindent \textit{Temperature-dependent isomeric shift.} The unsplit transition frequency described above shifts due to nonzero electronic wavefunction density at the nucleus~\cite{Greenwood1971}. The electron density changes with \textit{T} due to lattice expansion and thermal population of higher energy phonon modes. This subsequently rearranges the crystal electronic structure, which shifts the unsplit transition energy of the nuclear transition (Fig.~\ref{fig:setup}c). This effect is called the isomeric shift~\cite{Shirley1964}. The thorium-229 nuclear transition is therefore dependent on the chemical environment of the host crystal. It is typically assumed that the primary contribution to the isomeric shift arises from \textit{s} electrons due to their high wavefunction probability at the nucleus, although \textit{p}, \textit{d}, and \textit{f} electrons can contribute to the shift through screening effects~\cite{Shirley1964}. 

The \textit{T}-dependent isomeric shift is given by~\cite{Shirley1964},
\begin{equation}
    \delta \nu = \left( \frac{4 \pi Z e^2 R^2}{5h} \right) S'(Z)[\delta \psi ^2 (0)] \left[ \frac{\Delta R}{R} \right].
    \label{eq:chemicalShift}
\end{equation}

\noindent Here, $\nu$ is the unsplit transition frequency, \textit{Z} is the charge number of the nucleus, \textit{e} is the electron charge, \textit{R} is the nuclear radius, and $h$ is Planck's constant. $S'(Z)$ is a dimensionless “relativity factor” arising from relativistic corrections to electron charge density in heavy nuclei. In this work we use the value listed in Ref.~\cite{Shirley1964} of $S'(Z)$ = 11.68 for \textit{Z} = 90. The factor [$\delta$$\psi$$^2$(0)] is the change in the electron density at the nucleus (with units of inverse volume) as the crystal expands with \textit{T} increase. The value [$\Delta$R/R] is the nuclear factor containing the change in nuclear charge radius $\Delta$\textit{R} between the ground and isomeric states.

The reduction of the unsplit transition frequency from 150\,K to 293\,K indicates that the electronic charge density at the nucleus decreases when \textit{T} is increased. Using the value of $\Delta R^2$ = 0.0105(3)\,fm$^2$ from Ref.~\cite{Beeks2024} and the formula $\Delta R^2 = 2R\Delta R$, the change in electron density at the nucleus is [$\delta$$\psi$$^2$(0)]~=~--0.161(5)\,\AA$^{-3}$. The quoted uncertainty does not account for approximations in calculating the relativity factor $S'(Z)$.

\begin{figure}[t!]
    \includegraphics[width=\columnwidth]{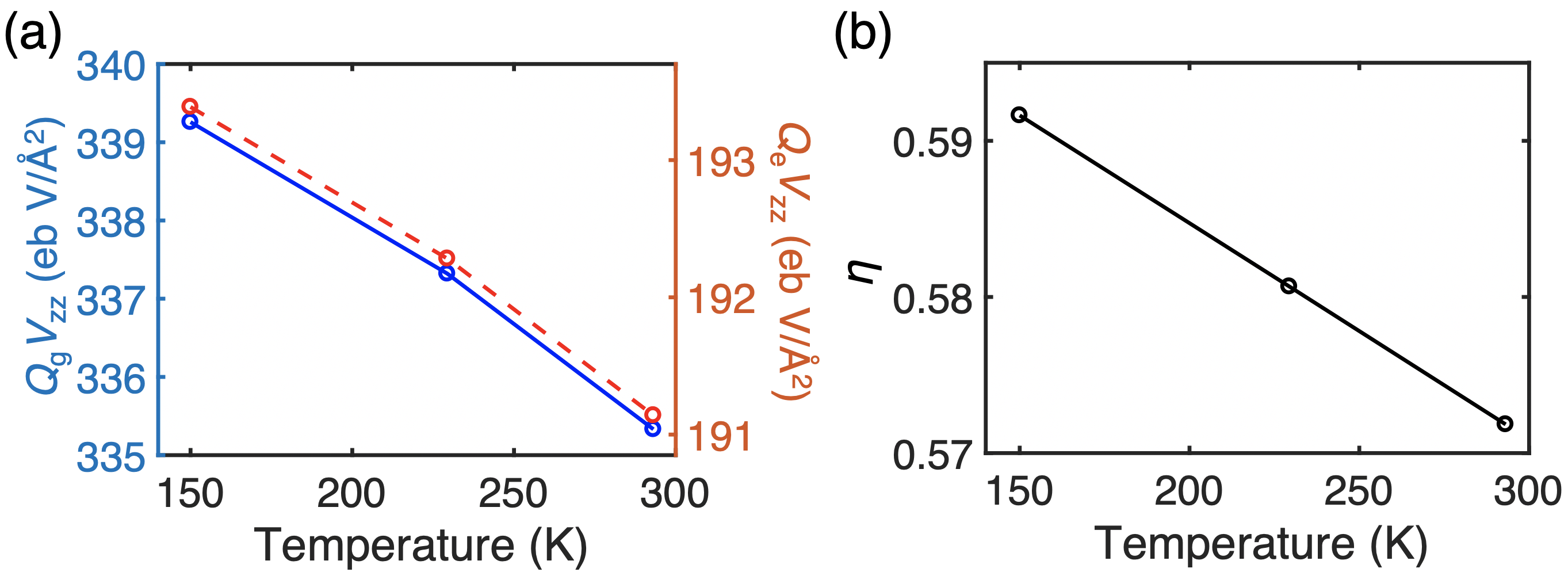}
    \caption{
        Electric quadrupole fitting parameters for each temperature. (a) The fit values for \Qg\Vzz{} and \Qe\Vzz{} are plotted along axes (blue, solid, left; red, dashed, right) with ranges that differ by the measured quadrupole ratio \Qe/\Qg. The electric field gradient decreases over the measured temperature range. (b) The fitted electric field gradient asymmetry $\eta$ (see Main Text) also decreases as the crystal temperature increases.}
    \label{fig:fit_plot}
\end{figure}

\begin{table*}[t]
\caption{\label{tab:fits}%
Fitted parameters at three different temperatures, corresponding to Fig \ref{fig:fit_plot}.}
\begin{tabular}{|c|cccc|}
\hline
Temperature (K) & \multicolumn{1}{c}{\Vzz\Qg~(eb V$/$\AA$^2$)} & \Vzz\Qe~(eb V$/$\AA$^2$) & $\eta$     & Isomeric shift (kHz) \\ \hline
150(1)             & 339.258(7)  & 193.387(5)                                                                          & 0.59163(5) & 0                      \\
229(1)           & 337.320(4)   & 192.282(3)                                                                          & 0.58067(3) & -409(4)                \\
293(1)           & 335.331(9)  & 191.140(5)                                                                          & 0.57184(5) & -776(3)                \\ \hline
\end{tabular}
\end{table*}
}

\noindent \textit{Temperature-dependent quadrupole splittings.} The nuclear quadrupole structure arises from the interaction of the EFG of the crystal and the electric quadrupole moment of the nucleus. The Hamiltonian is given by~\cite{Dunlap1985}
\begin{equation}
    H= \frac{QV_{zz}}{4I(2I-1)} \left[ 3I_{z}^{2} - \textbf{I}^{2} + \eta\left( I_{x}^{2} - I_{y}^{2} \right) \right].
    \label{eq:quadrupole}
\end{equation}

\noindent Here, \textit{Q} is the spectroscopic quadrupole moment in the laboratory frame with \Qg (\Qe) being the ground state (isomeric) quadrupole moment, \textit{I} is the total nuclear spin, and $\eta$ is the asymmetry parameter given by $\eta$~=~(\Vxx-\Vyy)/\Vzz. In this formalism, we assume the axis where the EFG is diagonalized, and \Vzz $>$ \Vxx $>$ \Vyy{} denote the \textit{x}, \textit{y}, and \textit{z} components of the EFG. Nonzero values for $\eta$ induce state mixing between the quadrupole levels and therefore alter transition strengths. Our analysis accounts for the frequency shifts but does not use line intensity.

We extract the fitting parameters in a similar manner to Ref.~\cite{Zhang2024}. We generate a distribution of sets of transition frequencies given by the uncertainties in Table ~\ref{tab:freqs}. We diagonalize the Hamiltonian in Eq.~\ref{eq:quadrupole} and fit each set of frequencies to the parameters $\eta$, \Qg\Vzz, and \Qe\Vzz. This yields a distribution of fit parameters, from which we take the average and standard deviation to extract the values and their uncertainty. We conduct these fits by both fixing the quadrupole ratio \Qe/\Qg{} at 0.57003(1) (taken from Ref.~\cite{Zhang2024}) for all temperatures and by letting the ratio be a free parameter. In each case, the resulting fit parameters agree within statistical uncertainty. In what follows, the fits from the floating quadrupole ratio are reported. 

Table~\ref{tab:fits} and Fig.~\ref{fig:fit_plot} show the fit parameters for the three measured temperatures. In Fig.~\ref{fig:fit_plot}a, the parameters \Qg\Vzz ~and \Qe\Vzz ~are superimposed with the ratio of axis scale given by the quadrupole ratio. The EFG \Vzz ~decreases by 1.2$\%$ when \textit{T} is increased from 150\,K to 293\,K. With an estimated quadrupole moment \Qg = 3.11(2)\,eb given by Ref.~\cite{Porsev2021}, the EFG values for 150\,K, 229\,K, and 293\,K are 109.1(7), 108.5(7), and 107.8(7)\,V/Å$^{2}$, respectively. Fig.~\ref{fig:fit_plot}b shows the changing asymmetry parameter $\eta$ plotted against \textit{T}. Here, $\eta$ also decreases with \textit{T} by 3.3$\%$ over the entire temperature range, corresponding to an anisotropic thermal expansion of the crystal lattice. Similar \textit{T}-dependent asymmetry parameters have also been observed in previous Mössbauer spectroscopy studies \cite{Nascimento1977,Weber2014}.


\noindent \textit{Discussion.} In solid state nuclear clocks, where local crystal environment causes shifts and splittings of the transition, the temperature dependence is a crucial systematic to characterize clock stability. Here, we measure four of the five transition lines within the electric quadrupole structure in thorium-doped calcium fluoride at three temperatures ranging over $\sim$150\,K. We find that the frequencies of three of the transitions shift over a range of a few MHz when heated from 150\,K to 293\,K, representing a fractional shift of 10$^{–12}$\,K$^{–1}$ to 10$^{–11}$\,K$^{–1}$, in line with the \textit{T}-dependent shifts predicted previously~\cite{Rellergert2010, Kazakov2012}. Our measurement uncertainty at the kHz level is likely limited by the $\sim$300\,kHz linewidth of the VUV frequency comb. We do not yet know the degree of inhomogeneous broadening due to the crystal environment, but this can be measured by further narrowing the comb linewidth. Regardless, this study shows that the transitions have a \textit{maximum} inhomogeneous linewidth on the order of $\sim$100\,kHz, which is lower than other condensed phase optical transitions that exhibit room temperature linewidths on the order of ~100 MHz or higher~\cite{Evans2016, Ngan2023, Ashley2024}.
 
We observe a \textit{T}-dependent isomeric shift of the nuclear transition. The unsplit transition frequency and electric quadrupole splittings both decrease as \textit{T} is increased, corresponding to a decrease in electronic charge density, EFG, and EFG asymmetry at the position of the nucleus. Upon heating, electron density is pulled away from the thorium nucleus from expansion of the positions of nuclei, likely via the highly electronegative fluoride anions in the crystal lattice and/or in charge compensation sites. These results can benchmark density functional theory coupled with molecular dynamics methods to understand the \textit{T}-dependent changes in crystal structure and their subsequent effect on electronic structure~\cite{Car1985}. Electronic structure calculations that precisely calculate the change in charge density could possibly be used in tandem with Eq.~\ref{eq:chemicalShift} to extract more precise nuclear parameters such as the change in the nuclear charge radius [$\Delta R/R$] between the ground and isomeric state. This study shows that changes in the electronic structure of solids can be sensitively probed using tabletop Mössbauer spectroscopy~\cite{Gutlich2012}. 

We observe one transition, \textit{m} = $\pm$5/2 → $\pm$3/2, that is much less sensitive to temperature shifts due to a fortuitous partial cancellation of the electric quadrupole splitting and isomeric shift. This transition changes by 62(6)\,kHz over the 143\,K range, whereas the other transitions each change by more than 1\,MHz. Additionally, it has the strongest Clebsch-Gordan coefficient among the electric quadrupole splittings~\cite{Kazakov2012}. This transition is therefore a promising candidate for future clock operation. Precision at the 10$^{-18}$ level of this transition requires a frequency stability of 2 mHz, which corresponds to a temperature stability on the order of 5 $\upmu$K, given the above temperature dependence. This differential temperature dependence of various lines also opens the possibility of setting up a co-thermometry to remove the temperature-dependent frequency shift. This level of temperature sensitivity has been demonstrated in multiple platforms~\cite{Weng2014, Reihani2022, Gong2024}. We note that between 150\,K and 229\,K, the transition only changes by 5(5)\,kHz, indicating a possible further reduction in temperature sensitivity over this range; more data are required to verify this trend. Further, material engineering of the sample such as introducing controlled mechanical strain may offer a tunable method to change the crystal properties, possibly leading to further cancellation of temperature shifts and a more stable clock transition. Other systematic shifts and broadening mechanisms include magnetic dipole broadening (estimated to be on the order of $\sim$400\,Hz) and second order Doppler effects (estimated shift at $\sim$1\,Hz/K with broadening on the order of $\sim$1\,Hz/K)~\cite{Rellergert2010,Kazakov2012}. Future studies should narrow the VUV laser linewidth~\cite{Benko2014} to fully characterize these potential systematic shifts.

\noindent \textit{Acknowledgments.} 
We thank Kim Hagen for technical assistance.
We also acknowledge funding support from the Army Research Office (W911NF2010182); Air Force Office of Scientific Research (FA9550-19-1-0148); National Science Foundation QLCI OMA-2016244; DOE quantum center of Quantum System Accelerator; National Science Foundation PHY-2317149; and National Institute of Standards and Technology.
Part of this work has been funded by the European Research Council (ERC) under the European Union’s Horizon 2020 research and innovation programme (Grant Agreement No. 856415) and the Austrian Science Fund (FWF) [Grant DOI: 10.55776/F1004, 10.55776/J4834, 10.55776/ PIN9526523]. The project 23FUN03 HIOC [Grant DOI: 10.13039/100019599] has received funding from the European Partnership on Metrology, co-financed from the European Union’s Horizon Europe Research and Innovation Program and by the Participating States. K.B. acknowledges support from the Schweizerischer Nationalfonds (SNF), fund 514788 “Wavefunction engineering for controlled nuclear decays."


\noindent \textbf{End Matter} 

\noindent \textit{Methods.} We measure the absolute frequency of the transition through direct VUV frequency comb spectroscopy, described in detail in previous publications~\cite{Zhang2024,Cingoz2012,Zhang2022}. Our setup is shown in Fig.~\ref{fig:setup}a. A \thor-doped calcium fluoride crystal (\thor:CaF$_2$ grown at TU Wien, see Refs.~\cite{Beeks2023,Beeks2024caf2} for details) is cooled in a vacuum chamber via an external dewar connected to a cold trap and a copper thermal link. The reported \textit{T} is measured using a sensor placed on the metal baseplate where the crystal is mounted. Measurements were performed at three temperatures: 150(1) K using liquid nitrogen, 229(1) K using a mixture of dry ice and methanol, and 293(1) K at room temperature. The temperature readout remained stable to within 0.5 K while measuring each temperature range. During the measurement time, there could be a slight steady state temperature gradient between the crystal temperature and the sensor. The sensor was not fully calibrated to the temperature within the vacuum chamber but had a consistent daily readout. With these factors in mind, we conservatively placed a 1 K uncertainty on the temperature, though the uncertainty is likely smaller.

We conduct scans by changing the comb repetition frequency, which shifts the VUV comb mode spacing. For each spectroscopic feature, the comb mode exciting the transition at 150 K was determined in our previous work~\cite{Zhang2024}. We expect that the \textit{T}-dependent shift is on the order of $\sim$1\,MHz or less~\cite{Rellergert2010,Kazakov2012}, which is much less than the comb mode spacing of 75 MHz (Fig.~\ref{fig:setup}b). Thus, the comb mode number exciting a given transition remains the same for all temperatures. We find that the measured transition lines are in agreement with expected electric quadrupole splittings (see Main Text), further indicating that our assumption is valid. The center repetition frequency of each line is determined using previously described methods~\cite{Zhang2024}. The absolute frequency for each feature is determined from the frequency comb equation,
\begin{equation}
    \nu_\mathrm{Th}=Nf_\mathrm{rep}+7f_\mathrm{CEO},
    \label{eq:combEquation}
\end{equation}

\noindent where $N$ is the integer comb mode number ($\sim$2.7~$\times$~10$^{7}$), $f_{\rm rep}$ is the comb repetition frequency, and $f_\mathrm{CEO}$, the carrier envelope offset frequency, is –8\,MHz in this study. The uncertainty of each transition frequency is determined by multiplying the center frequency uncertainty in \textit{f}$_{\mathrm{rep}}$ by the comb mode number.\\

\bibliography{ref}

\end{document}